\documentclass[aps,twocolumn,prb,floatfix,superscriptaddress,showpacs,lengthcheck,hyperref,citeautoscript]{revtex4-1}
\usepackage{amsmath}
\usepackage{amssymb}
\usepackage{graphicx}
\usepackage{epstopdf}
\usepackage{bm}
\usepackage{color}
\usepackage{times}
\usepackage{MnSymbol}

\newcommand{\ve}{\varepsilon}

\newcommand{\bea}{\begin{eqnarray}}
\newcommand{\eea}{\end{eqnarray}}
\newcommand{\be}{\begin{equation}}
\newcommand{\ee}{\end{equation}}

\newcommand{\Ht}{\tilde{\mathcal{H}}}

\newcommand{\ci}{\mathrm{i}}
\newcommand{\la}{\langle}
\newcommand{\ra}{\rangle}
\newcommand{\ket}[1]{| #1 \rangle}
\newcommand{\bra}[1]{\langle #1 |}
\newcommand{\bep}{\bar{\varepsilon}}

\begin{document}

\title{Irradiated graphene as a tunable Floquet topological insulator}
\author{Gonzalo Usaj}
\affiliation{Centro At{\'{o}}mico Bariloche and Instituto Balseiro,
Comisi\'on Nacional de Energ\'{\i}a At\'omica, 8400 Bariloche, Argentina}
\affiliation{Consejo Nacional de Investigaciones Cient\'{\i}ficas y T\'ecnicas (CONICET), Argentina}
\author{P. M. Perez-Piskunow}
\affiliation{Instituto de F\'{\i}sica Enrique Gaviola (CONICET) and FaMAF, Universidad Nacional de C\'ordoba, Argentina}
\author{L. E. F. Foa Torres }
\affiliation{Instituto de F\'{\i}sica Enrique Gaviola (CONICET) and FaMAF, Universidad Nacional de C\'ordoba, Argentina}
\author{C. A. Balseiro}
\affiliation{Centro At{\'{o}}mico Bariloche and Instituto Balseiro,
Comisi\'on Nacional de Energ\'{\i}a At\'omica, 8400 Bariloche, Argentina}
\affiliation{Consejo Nacional de Investigaciones Cient\'{\i}ficas y T\'ecnicas (CONICET), Argentina}
\begin{abstract}
In the presence of a circularly polarized mid-infrared radiation graphene develops dynamical band gaps in its quasi-energy band structure and becomes a Floquet insulator. Here we analyze how topologically protected edge states arise inside these gaps in the presence of an edge. Our results show that the gap appearing at $\hbar\Omega/2$, where $\hbar \Omega$ is the photon energy,  is bridged by two chiral edge states whose propagation direction is set by the direction of the polarization of the radiation field. Therefore, both the propagation direction and the energy window where the states appear can be controlled externally. We present both analytical and numerical calculations that fully characterize these states. This is complemented by simple topological arguments that account for them and by numerical calculations for the case of semi-infinite sample, thereby eliminating finite size effects. 
\end{abstract}
\date{\today}
\maketitle

\section{Introduction}

Graphene is an extraordinary material with unusual electrical\cite{Novoselov2005a,CastroNeto2009}, mechanical, thermal\cite{Balandin2008} and optical properties\cite{Bonaccorso2010}. However, probably one of the most desirable but still missing property is the presence of topologically protected states such as those found in topological insulators (TIs).\cite{Kane2005,Koenig2007,Hasan2010} Although one of the pioneering works that propelled the whole field of TIs was based on Dirac fermions in graphene,\cite{Kane2005} the spin-orbit coupling turns out to be too weak for a topological phase to be observed. Since the number of known materials behaving as TIs is limited, bringing these properties to carbon-based materials\cite{FoaTorres2014} with the addition of a fully-fledged tunability may enormously expand their prospects.

Manipulating the electronic structure of matter by coupling electrons and photons into entangled states has been a subject of intense activity for many years. In the present context, harnessing light-matter interaction\cite{Hartmann2014,Glazov2014} may offer a wealth of novel phenomena,\cite{Karch2011,Dora2012,Guo2013,Tielrooij2013} such as Floquet-Majorana modes,\cite{Reynoso2013,Kundu2013,Thakurathi2013,Sato2014} or allow the manipulation of Dirac points \cite{Rodriguez-Lopez2013,Scholz2013}. Furthermore, time-dependent driving may provide for unexpected ways of turning a normal material into a special topological insulator,\cite{Oka2009,Lindner2011,Kitagawa2011,SuarezMorell2012} also called Floquet Topological Insulator (FTI) \cite{Lindner2011,Gu2011,Gomez-Leon2013,Rudner2013}. The interest in these novel non-equilibrium phases of topological order is increasing\cite{Cayssol2013,Zhou2014,TenenbaumKatan2013a,TenenbaumKatan2013} not only in condensed matter \cite{Sentef2014,Kundu2014,Dehghani2014} and cold atoms\cite{Atala2013,Goldman2014,Choudhury2014} but also from a more general point of view as a new classification may be needed.\cite{Rudner2013,Ho2014,Asboth2014}

First, one would need to open up a gap in the material's bulk and then one should check for the presence of topological edge states. Laser-induced bandgaps were predicted to occur for Dirac fermions under a circularly polarized laser \cite{Oka2009,Kibis2010,Savelev2011,Zhou2011,Iurov2012} in a feasible range of parameters (mid-infrared range (MIR)) and being polarization tunable \cite{Calvo2011,Calvo2012a}. Now, two recent experiments add new thrill to this area from different perspectives: The first is the realization of a FTI in a hexagonal lattice crafted in a photonic crystal \cite{Rechtsman2013}; the second one is the observation of a polarization tunable band structure at the surface of a topological insulator through ARPES.\cite{Wang2013} This last experiment showed the emergence of the dynamical gaps by using circularly polarized light in the MIR.

\begin{figure}[t]
\centering\includegraphics[width=\columnwidth]{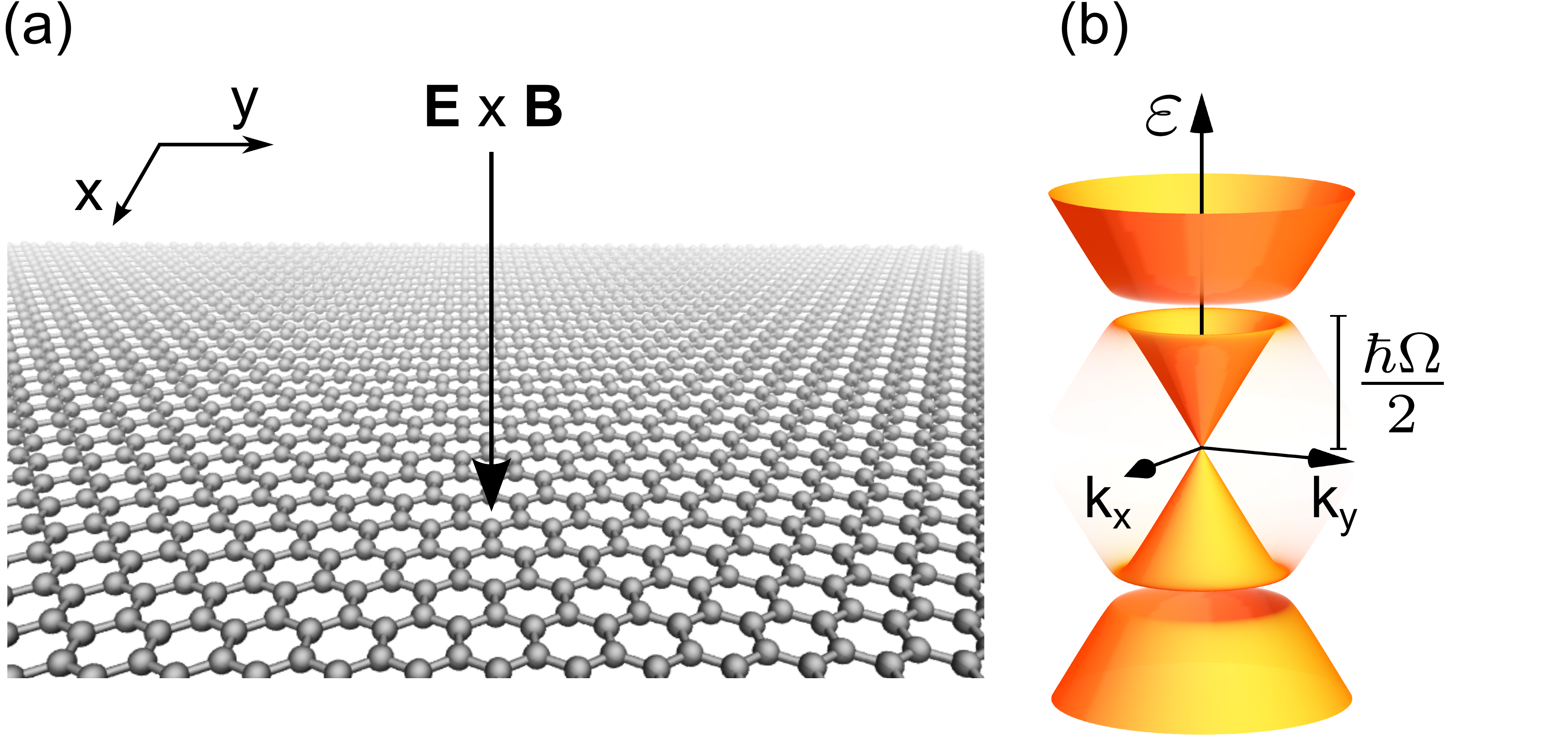}
\caption{(color online) (a) Scheme of a bulk graphene sheet being illuminated by a laser (perpendicularly to the graphene plane). (b) Scheme showing how the Dirac cone is being modified by the laser. The opening of a dynamical gaps at $\pm \hbar\Omega/2$ is evident. The bands shown in this scheme are weighted on the $m=0$ Floquet channel, \textit{i.e.} these bands are the ones contributing to the dc density of states.}
\label{scheme}
\end{figure}

Here we extend on our recent proposal for achieving Floquet chiral edge states in graphene through laser illumination \cite{Perez-Piskunow2014}. Before, we showed that a carefully tuned circularly polarized laser may introduce a bulk dynamical bandgap at half the photon energy\cite{Calvo2011} (a scheme of the bulk dispersion is shown in Fig. \ref{scheme}) while keeping propagating states through the edges of a zigzag ribbon \cite{Perez-Piskunow2014}. Interestingly, these Floquet edge states turn out to be chiral. Many important fundamental and technical aspects however remained. The search for Floquet topological states may benefit from more accurate and diverse experimental proposals.\cite{Fregoso2014} Here we provide a detailed analytical derivation which is complemented by a simple discussion of the topological character of the bulk bands. The topological analysis provides hints for predicting the fate of these states when disorder is included. Moreover, the role of different types of ribbon terminations and the band structure of a \textit{semi-infinite} sample are also addressed numerically. The latter eliminates finite-size effects and allows a direct verification of the strengths and limits of the topological analysis.

\section{Irradiated graphene: Bulk properties}
In the presence of  electromagnetic radiation, the electronic states of bulk graphene close to the Dirac point are described by the following time-dependent Hamiltonian
\begin{equation}
\hat{\cal{H}}(t)=v_F\,\bm{\sigma}\cdot\left[\bm{p}+
\frac{e}{c}\bm{A}(t)\right]\,, 
\label{eq_laser-graphene-Hamiltonian} 
\end{equation}
where $v_{F}\simeq 10^{6} m/s$ denotes the Fermi velocity, $\bm{\sigma}=(\sigma_x,\sigma_y)$ the Pauli matrices describing the pseudo-spin degree of freedom, $e$ the absolute value of the electron charge, $c$ the speed of light and $\bm{A}(t)=\Re\left\{\bm{A}_0e^{\mathrm{i}\Omega t}\right\}$ the vector potential of the electromagnetic field (incident perpendicularly to the graphene sheet). We consider the circularly polarized case, where $\bm{A}_{0}=A_{0} (\hat{\bm{x}}+\ci\hat{\bm{y}})$, and assume that the laser spot is much larger than the system size  in order to neglect any spatial dependence. The choice of circular rather than linear polarization is a subtle but important one: In contrast to linear polarization, circular polarization breaks time-reversal symmetry and allows for non-trivial topological properties \cite{Oka2009,Rudner2013} and Floquet chiral edge states.\cite{Perez-Piskunow2014}

\subsection{Floquet theory}
For what follows, it is instructive to briefly introduce the basic ideas of the Floquet formalism\cite{Shirley1965,Sambe1973} used to deal with time dependent periodic Hamiltonians (for more extensive general reviews we refer to Refs. [\onlinecite{Grifoni1998}] and [\onlinecite{Kohler2005}]; in the context of graphene a shorter account on this technique is given in  Ref.~[\onlinecite{FoaTorres2014}]). 

Floquet theorem guarantees the existence of a set of solutions of the time-dependent Schr\"odinger equation of the form $\ket{\psi_{\alpha}(t)}=\exp(-\ci\varepsilon_{\alpha}t/\hbar)\ket{\phi_{\alpha}(t)}$ where $\ket{\phi_{\alpha}(t)}$ has the same time-periodicity as the Hamiltonian, $\ket{\phi_{\alpha}(t+T)}=\ket{\phi_\alpha(t)}$ with $T=2\pi/\Omega$.\cite{Shirley1965,Grifoni1998,notes3} The Floquet states $\ket{\phi_{\alpha}}$ are the solutions of the equation
\begin{equation}
\hat{\cal{H}}_F\ket{\phi_{\alpha}(t)} =\varepsilon_{\alpha}\ket{\phi_{\alpha}(t)}\,,
\label{floquetEq}
\end{equation}
where $\hat{\cal{H}}_F=\hat{\cal{H}}-\ci\hbar \frac{\partial}{\partial t}$ is the Floquet Hamiltonian and $\varepsilon_{\alpha}$ the quasi-energy. 

Using the fact that the Floquet eigenfunctions are periodic in time, it is customary to introduce an extended $\cal{R}\otimes \cal{T}$ space, where $\cal{R}$ is the usual Hilbert space and $\cal{T}$ is the space of periodic functions with period $T$.
In this space, also called Floquet or Sambe space,\cite{Sambe1973} we can define the inner product
\begin{equation}
\langle\langle \phi_{\alpha}(t)|\phi_{\beta}(t)\rangle\rangle=\frac{1}{T}\int_0^T \langle \phi_{\alpha}(t)|\phi_{\beta}(t)\rangle\, dt\,,
\label{inner}
\end{equation} 
from which it is easy to show that $\hat{\cal{H}}_F$ is an Hermitian operator. This implies that $\langle\langle \phi_\alpha|\phi_\beta\rangle\rangle=\delta_{\alpha\beta}$ for any pair of eigenvectors. Yet, it is important to note that  while $\ket{\phi^{(n)}_{\alpha}}=e^{\ci n\Omega t}\ket{\phi_{\alpha}}$, which is also a solution of Eq.~(\ref{floquetEq}) with quasi-energy  $\varepsilon_\alpha^{(n)}=\varepsilon_\alpha+n\hbar\omega$ for an arbitrary integer $n$,  and $\ket{\phi_{\alpha}}$  are orthogonal in $\cal{R}\otimes \cal{T}$ (for $n\neq0$),
\begin{equation}
\langle\langle \phi_{\alpha}(t)|\phi_{\alpha}^{(n)}(t)\rangle\rangle=\delta_{n0}\,.
\label{orto1}
\end{equation}
they correspond to the same physical state. Namely, 
\begin{equation}
\ket{\psi_{\alpha}(t)}=e^{-\ci\varepsilon_{\alpha}t/\hbar}\ket{\phi_{\alpha}(t)}=e^{-\ci\varepsilon^{(n)}_{\alpha}t/\hbar}\ket{\phi^{(n)}_{\alpha}}\,.
\end{equation}
Therefore, all non-equivalent physical states are restricted to a quasi-energy window of $\hbar\Omega$ around any given quasi-energy $\varepsilon_\alpha$ (the so-called Floquet zone (FZ))---of course, we can still use an `extended FZ' picture as in the more usual case of Bloch band states; we use that picture in the following sections as it is better suit for a physical interpretation of the results.

The Floquet eigenfunctions, when restricted to a given FZ, satisfy the following orthogonality and closure relations in $\cal{R}$ for a \textit{fixed} time $t$,
\begin{eqnarray}
\label{orto2}
\langle \phi_{\alpha}(t)|\phi_{\beta}(t)\rangle&=&\delta_{\alpha\beta}\,,\\
\sum_\alpha \ket{\phi_{\alpha}(t)}\bra{\phi_{\alpha}(t)}&=&\bm{I}\,.
\end{eqnarray}
A convenient basis of $\cal{R}\otimes \cal{T}$ can be built from the product of an arbitrary basis of $\cal{R}$ (the eigenfunctions of the time-independent part of the Hamiltonian, for instance) and the set of orthonormal functions $e^{\ci m\Omega t}$, with $m=0,\pm1,\pm2, ...$ that span $\cal{T}$. Then, 
\begin{equation}
\label{fourier}
\ket{\phi_{\alpha}(t)}=\sum_{m=-\infty}^\infty \ket{u_m^\alpha}\,e^{\ci m\Omega t}\,,
\end{equation}
or, in a vector notation in  $\cal{R}\otimes \cal{T}$,
\begin{equation}
\ket{\phi_{\alpha}}=\{\cdots,\ket{u_1^\alpha},\ket{u_0^\alpha},\ket{u_{-1}^\alpha},\cdots\}^\mathrm{T}\,.
\label{vector}
\end{equation}
Written in this basis, $\hat{\cal{H}}_F $ is a time-independent infinite matrix operator $\tilde{\mathcal{H}}_F^\infty$ with Floquet replicas shifted by a diagonal term $m\hbar\Omega$  and coupled by the radiation field with the condition, for pure harmonic potentials, $\Delta m=\pm1$
\begin{widetext}
\begin{equation}
\tilde{\mathcal{H}}_F^\infty=\left(
\begin{array}{cccccc}
\ddots&\vdots&\vdots&\vdots&\vdots&\udots\\
\cdots&v_F\bm{p}\cdot\bm{\sigma}+2\hbar\Omega \bm{I}&\frac{v_Fe}{2c}A_0 \sigma_- & \bm{0}  &  \bm{0}  &\cdots\\
\cdots&\frac{v_Fe}{2c}A_0\sigma_+ &v_F \bm{p}\cdot\bm{\sigma}+\hbar\Omega \bm{I}& \frac{v_Fe}{2c}A_0 \sigma_-  &  \bm{0} &\cdots  \\
\cdots&  \bm{0} & \frac{v_Fe}{2c}A_0 \sigma_+& v_F \bm{p}\cdot\bm{\sigma} &\frac{v_Fe}{2c} A_0 \sigma_-  &\cdots\\
\cdots&  \bm{0} &  \bm{0} & \frac{v_Fe}{2c}A_0 \sigma_+  & v_F\bm{p}\cdot\bm{\sigma} -\hbar\Omega \bm{I}&\cdots  \\
\udots& \vdots&\vdots&\vdots&\vdots&\ddots\\
\end{array}
\right)\,.
\label{matrix_inf}
\end{equation}
\end{widetext}
Here $\sigma_\pm=(\sigma_x\pm\ci\sigma_y)$. Thus, Eq.(\ref{floquetEq}) becomes a time-independent eigenvalue problem. 

Since we are interested on the Floquet spectrum around the dynamical gap, that is $\varepsilon\sim\hbar\Omega/2$, we restrict the Floquet Hamiltonian to the $m=0$ and $m=1$ subspaces (or replicas) for the analytical calculations---the numerical results can retain a larger number ($N_\mathrm{FR}$) of replicas. As we will show, this restriction is enough to get the main features of the energy dispersion and the Floquet states.
The reduced Floquet Hamiltonian describing states close to the $K$ point of the graphene's Brillouin Zone then corresponds to the central blocks of Eq.~(\ref{matrix_inf}),
\begin{equation}
\tilde{\mathcal{H}}_F=\left(
\begin{array}{cc}
v_F \bm{p}\cdot\bm{\sigma}+\hbar\Omega \bm{I}& \frac{v_Fe}{2c}A_0 \sigma_- \\
\frac{v_Fe}{2c}A_0 \sigma_+& v_F \bm{p}\cdot\bm{\sigma}
\end{array}
\right)\,,
\label{matrix_p}
\end{equation}
In the notation of Eq. (\ref{vector}), the Floquet equation $\tilde{\mathcal{H}}_F\ket{\phi}=\varepsilon \ket{\phi}$ involves finding a four-component wave function 
\begin{equation}
\phi(\bm{r})=\{[u_{1A}(\bm{r}),u_{1B}(\bm{r})],[u_{0A}(\bm{r}),u_{0B}(\bm{r})]\}^\mathrm{T},
\end{equation}
where each component $u_{mi}(\bm{r})$ refers to the $m=0,1$ subspace and the $i=A,B$ to the lattice site---we include the square brackets in the notation to emphasize the spinor character of the wave-function on each replica.
\subsection{The bulk states} 
The Floquet states in a \textit{bulk} graphene sheet have been discussed in several works for both linear \cite{Syzranov2008} and circular polarization.\cite{Oka2009} They are the starting point for our study the formation of laser-induced bandgaps and the emergence of non-trivial topological properties and, for the sake of completeness, we present here a simple derivation. 

Due to the translational invariance the wave function takes the form 
\begin{equation}
\phi_{\bm{k}\alpha}(\bm{r})=\mathrm{e}^{\ci \bm{k}\cdot\bm{r}}\{[u^{\bm{k}\alpha}_{1A},u^{\bm{k}\alpha}_{1B}],[u^{\bm{k}\alpha}_{0A},u^{\bm{k}\alpha}_{0B}]\}^\mathrm{T}\,, 
\label{wf-bulk}
\end{equation}
where the index $\alpha$ denotes the four solutions of the Floquet Hamiltonian 
\begin{equation}
\tilde{\mathcal{H}}_F=\left(
\begin{array}{cccc}
\frac{\hbar\Omega}{2} & \hbar v_F k_- & 0 & 0\\
\hbar v_F k_+&\frac{\hbar\Omega}{2} &  \frac{v_Fe}{c}A_0 & 0 \\
0& \frac{v_Fe}{c}A_0 &-\frac{\hbar\Omega}{2} &\hbar v_F k_- \\
0&0 &\hbar v_F k_+ & -\frac{\hbar\Omega}{2}
\end{array}
\right)+\frac{\hbar\Omega}{2} \bm{I}
\label{matrix_k}
\end{equation}
with  energies ${\varepsilon_{\bm{k}\alpha}}$ and $k_{\pm}=k_x\pm \ci k_y$. For $A_{0}=0$ the Hamiltonian $\tilde{\mathcal{H}}_F$ has four eigenenergies: $\pm \hbar v_F  k$ and  $\hbar\Omega \pm \hbar v_F k$. Two of these eigenstates, $\hbar\Omega-\hbar v_F k$ and $\hbar v_F k$ become degenerate at $k=k_0=\Omega/2v_F$ where the quasi-energy value is $ \hbar\Omega/2$. A finite amplitude $A_{0}$ of the radiation mixes these two states generating an anti-crossing and opening a gap.\cite{notes4}
By introducing Eq. (\ref{wf-bulk}) into the Floquet equation, one can find the $u^{\bm{k}\alpha}_{mi}$ coefficients. In this case, however, it is convenient to further reduce the problem by solving the eigenvalue equation in the subspace of the two degenerate branches.
The Floquet quasi-energies of these branches near the degeneracy point are then given by 
\begin{equation}
\varepsilon_{k\pm}=\frac{\hbar\Omega}{2}(1+\mu_\pm)\,,\qquad \mu_\pm=\pm  \sqrt{\left(1-\frac{k}{k_0}\right)^2+\eta^2}\,,
\label{autoval}
\end{equation} 
where 
\begin{equation}
\eta=\frac{e v_F A_0}{c\hbar\Omega}
\end{equation}
is the dimensionless parameter controlling the transition from the weak to the strong coupling regime---we will always consider the case $\eta\ll1$ so the perturbative approach remains valid (the strong coupling regime was considered recently for a linearly polarized laser \cite{Syzranov2013}). 
The dynamical gap is $\eta \hbar\Omega=ev_FA_0/c$. The resulting dispersion of the Floquet quasi-energies is shown in Figure \ref{scheme-Chern}.  

Finally, the time-dependent solutions of the Schr\"odinger equation are
\begin{eqnarray}
\label{auto}
\psi_{\bm{k}\pm}(\bm{r},t)&=&\mathrm{e}^{-\ci\varepsilon_{k\pm}t/\hbar} \mathrm{e}^{\ci \bm{k}\cdot\bm{r}} \frac{1}{ \sqrt{2\mathcal{A}}} \\
\nonumber
&&\times 
\left[-\sin{(\varphi_{k}^{\pm}/2)}\,\binom{\mathrm{e}^{\ci\theta_{\bm{k}}}}{-1}\,\mathrm{e}^{\ci\Omega t}+\cos{(\varphi_{k}^{\pm}/2)}\,\binom{\mathrm{e}^{\ci\theta_{\bm{k}}}}{1}\right]\,.
\end{eqnarray}
Here $\mathcal{A}$ is the area of the graphene sheet, $\theta_{\bm{k}}$ is the angle formed by $\bm{k}$ and the $x$-axis and 
\begin{equation}
\tan{\varphi_{k}^{\pm}}=\eta \frac{k_0}{k-k_0}\,.
\end{equation}
The instantaneous expectation value of the velocity operator, $\bm{v}=v_F \bm{\sigma}$, evaluated in these states are
\begin{eqnarray}
\langle v_\parallel\rangle_{\bm{k}\pm}&=&  v_F \cos{\varphi_{k}^{\pm}}=v_F \frac{k_0-k}{k_0\,\mu_\pm}  \\
\nonumber
\la v_\perp\ra_{\bm{k}\pm}&=&v_F \sin{\varphi_{k}^{\pm}}\, \sin{\Omega t}=-v_F \frac{\eta}{\mu_\pm}\sin{\Omega t}\,
\end{eqnarray}  
with $v_\parallel$ and $v_\perp$ the velocity components parallel and perpendicular to the wave-vector $\bm{k}$, respectively. The time \textit{averaged} velocity is
\begin{eqnarray}
\nonumber
\la\langle v_\parallel\ra\rangle_{\bm{k}\pm}&=&  v_F \cos{\varphi_{k}^{\pm}}  \\
\la\la v_\perp\ra\ra_{\bm{k}\pm}&=&0\,.
\end{eqnarray}  
One can verify that $\la\la \bm{v} \ra\ra_{\bm{k}\pm}=(1/\hbar)\nabla_{\bm{k}}\varepsilon_{k\pm}$
 as expected from the Hellmann-Feynman theorem.\cite{Sambe1973} The eigenstates in Eq.~(\ref{auto}) then propagate (on average) in the direction of the wave-vector $\bm{k}$.
One can also verify that $\la \sigma_z\ra_{\bm{k}\pm}=\sin{ \varphi_{k}^{\pm}}\, \cos{\Omega t}$, so that the pseudo-spin is precessing around the $\hat{\bm{k}}$ axis with frequency $\Omega$, 
\be
\la \bm{\sigma}\ra_{\bm{k}\pm}= \cos\varphi_k^\pm\,  \hat{\bm{k}}+\sin{ \varphi_{k}^{\pm}}\, (\sin{\Omega t}\, \hat{\bm{z}}\times\hat{\bm{k}}+ \cos{\Omega t}\, \hat{\bm{z}})
\ee
The amplitude of the oscillation is maximum at $k=k_0$ where the states do not propagate ($\la\la v_\parallel\ra\ra_{\bm{k}\pm}=0$).


\subsection{Topological character of the Floquet bands}

While the description of the topological character of the energy bands for a time-independent system is a mature field, driven systems started to be discussed much more recently in this context.\cite{Oka2009,Lindner2011,Kitagawa2010,Kitagawa2011,Rudner2013} Here, we present a simple analysis highlighting the main features of interest for our discussion of Floquet chiral edge states. This analysis of the bulk properties allows us to infer the existence of robust edge states as the ones obtained analytically and numerically in the following sections.

To calculate the number of states inside a given Floquet gap one needs to look at the Chern numbers of the entire Floquet spectrum.\cite{Rudner2013} The Chern number of each Floquet band ($C_n$) gives the difference between the number of chiral states above and below each it,\cite{Hasan2010,Rudner2013} while the sum of all the Chern numbers below a given band gives the number of chiral states above it. A proper calculation of $C_n$ requires, in principle, to take into account all replicas, or at least the $\mathcal{O}(D/\hbar\Omega)$ replicas that overlap in the region of the gap of interest---here $D$ is the graphene's bandwidth---since only in that case the Floquet spectrum is actually gapped. The main contribution to $C_n$ comes from the region in $k$-space where anticrossings between replicas occur (see appendix). While in a time-independent problem there is no distinction between the contribution to $C_n$ coming from different regions in $k$-space,  we argue in the following that in the Floquet space there is a hierarchy of contributions, and thus a hierarchy of edge states. This hierarchy is based on the weight of the Floquet band on a given subspace, say the one with $m=0$, which is determined by the parameter $\eta$. The reason for this is that the calculation of the  dc properties of the system, such as the time averaged density of states,  imply a partial o a total projection on one replica.

\begin{figure}[tbp]
\centering\includegraphics[width=\columnwidth]{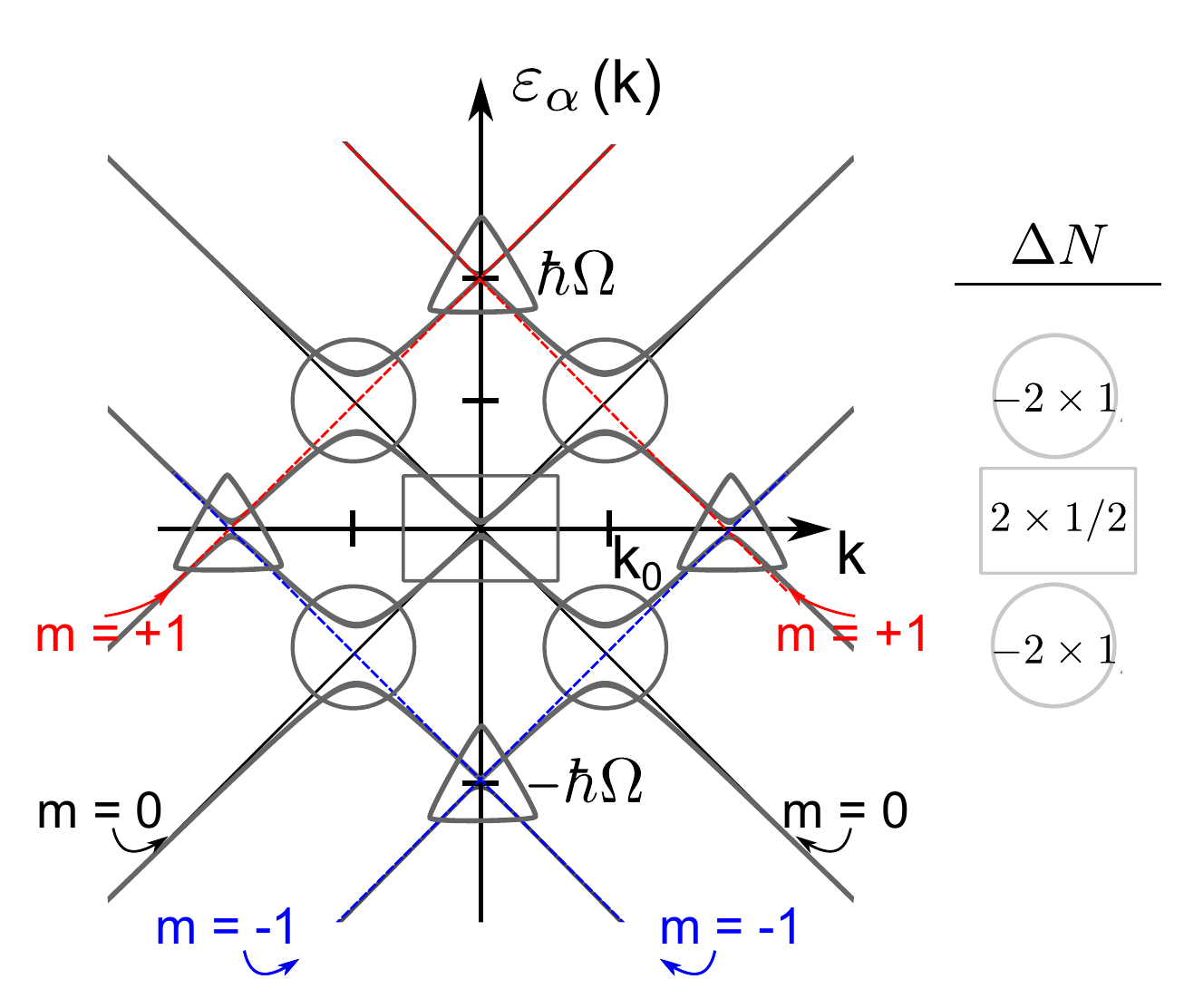}
\caption{(color online) Scheme of the Floquet bands with $m=-1,0,1$ as used for the calculation of the number of chiral edge states hosted within each gap. The table on the right indicates the number $\Delta N$ of chiral edges states at each crossing of the $m=0$ replica. }
\label{scheme-Chern}
\end{figure}

We start by truncating the Floquet Hamiltonian and consider the Floquet channels with $m=-1,0,1$. Then, the unperturbed spectrum projected on a given $\bm{k}$ direction looks like the one represented in Fig.~\ref{scheme-Chern}. Switching on the radiation opens bandgaps at the crossings located at the Dirac point and $\pm \hbar\Omega/2$. To infer the topological properties of these bands one could either do a numerical calculation of the Chern numbers for the full band structure (in the tight binding model) or an approximate calculation  as outlined below. 

For the approximate calculation one needs (see appendix for more details): \textit{(i)} to isolate each crossing where a bandgap opens;  \textit{(ii)} to obtain an effective Hamiltonian at each of those points (a $2\times2$ matrix of the form $\mathcal{H}^{\mathrm{eff}}_F(\bm{k},\nu)=\bm{h}_\nu(\bm{k})\cdot\bm{\sigma}$, $\nu=\pm 1$ the valley index); and  \textit{(iii)} to compute from it the contribution to the Chern number at each crossing (and sum over the two valleys), assuming that the associated Berry curvature decays fast enough away from them (similarly to what is done with bilayer graphene where one defines a valley Chern number\cite{Zhang2013}).

Let us start analyzing what happens to the Dirac point ($k\sim0$) in the $m=0$ replica (this region is marked with a rectangle in Fig. \ref{scheme-Chern}). A virtual photon process (absorption and then re-emission of one photon and viceversa) originates a gap,\cite{Oka2009,Calvo2011} which is of second order in the electron-photon coupling. In the large frequency limit, the effective Hamiltonian has $\bm{h}_\nu(\bm{k})=\hbar v_F (k_x,\nu k_y,\nu\,  \eta\, eA_0/\hbar c)$. This effective Hamiltonian describes the stroboscopic evolution of the system at each period $T=2\pi/\Omega$, just as if we had a time-independent system.\cite{Kitagawa2011} The contribution to the Chern number from each valley is $1/2$ (taking the limit $\eta\rightarrow 0$ at the end of the calculation) and, since they have the same sign, we get a total of $1$. Note that to get an integer number one needs to sum up the contributions from each valley, just as in Haldane's model\cite{Haldane1988} but this time in Floquet space\cite{Sentef2014}. Also, it should be kept in mind that this is a contribution of $+1$ to the Chern number of the Floquet band that is right below zero quasienergy. The band just above zero gets a contribution of $-1$.

The calculation  around $\hbar\Omega/2$ is more subtle since it involves a first order process in $\eta$ (circle in Fig. \ref{scheme-Chern}). We start by considering the truncated Floquet Hamiltonian of Eq. (\ref{matrix_k}). As before, to simplify the analysis even more it is convenient to consider only the subspace of the two degenerate branches with $m=0$ and $m=1$. The effective Floquet Hamiltonian has 
$\bm{h}(\bm{k})=\hbar v_F [(k-k_0)\,\hat{\bm{k}}-k_0\eta\,\hat{\bm{\zeta}}_{\bm{k}}]$, with $\hat{\bm{\zeta}}_{\bm{k}}=\sin\theta_{\bm{k}}\,\hat{\bm{\theta}}_{\bm{k}}+\cos\theta_{\bm{k}}\,\hat{\bm{z}}$,  
which gives a contribution  of $-2$ ($-1$ for each valley) to the Chern number of the Floquet band below $\hbar\Omega/2$. Adding this to the contribution coming from the region around $k\sim0$ we get a total contribution to the Chern number arising from these anticrossings of $-3$. We conclude that there should be a difference of $3$ in the chirality of the edge states\cite{Hasan2010,Rudner2013} appearing at the dynamical and the Dirac point gaps. Extending this procedure to all the Floquet bands in Fig.  \ref{scheme-Chern}  we conclude that $2$ \textit{edge states are expected to emerge at the dynamical gap (twice those at the smaller gap at the Dirac cone) with an $\mathcal{O}(1)$ weight on the $m=0$ subspace}.

Notice that the different signs of these contributions to the Chern number implies that the propagation direction of the associated edge states is also the opposite. This can also be appreciated in Fig. 7 where the dispersion weighted on the $m=0$ channel for a semi-infinite graphene sheet is shown, one distinguishes 2 states propagating to the left at the dynamical gap and one to the right close to the Dirac point. Given that  the dynamical gaps are linear in the laser strength, they are the best candidates for an experimental observation (indeed, the recent observation at the surface of a topological insulator\cite{Wang2013} highlights the dynamical gaps). 

A more careful inspection of Fig. \ref{scheme-Chern} shows that there are also other crossings taking place at zero energy and at $\hbar\Omega$: the ones marked with triangles  in Fig. \ref{scheme-Chern}. The situation in these cases is similar: laser-induced bandgaps emerge close to those points and turn out to host Floquet chiral edge states. But this is not the whole picture as the number of crossing points with zero-energy grows with the number of replicas considered when truncating the Floquet Hamiltonian---other appears at $\hbar\Omega/2$ if more replicas are included. Although this may seem irrelevant since those gaps turn out to be smaller and smaller (higher order in the radiation strength), an important question is whether this reduction of the gap in the overall quasi-energy spectrum effectively limits the range where topological properties are expected, and equally important, if it somehow weakens the topological protection. We argue that those ever smaller anticrossings do not give significant contribution to the time-averaged quantities, provided that the electron-radiation coupling is small, $\eta\ll1$. The main point is that those higher order states have a parametrically smaller weight (of order $\eta^{\delta m}$) on the $m=0$ channel and therefore do not contribute to the time-averaged density of states---$\delta m$ is the difference in the Floquet index of the two coupled replicas that leads to the high order gap.  Indeed, this can be appreciated (for the $\hbar\Omega/2$ gap) in Fig. 8, where we show a very fine detail close to the dynamical gap---a more detailed discussion is given in Sec.\ref{semi-infinite}. 
We therefore propose to use the $m=0$ Floquet-projected Chern number for our purposes.

In the next section we will pursue a different path to explicitly determine these states, their propagation velocity and decay length.


\section{Floquet topological states in zigzag edges: Analytical solution}
In this section, we present an analytical solution for the edge states near the dynamical gap by retaining only the $m=0$ and $m=1$ subspaces. While some particular cases of this solution (see Eqs. (\ref{ana-wave1}) and (\ref{ana-wave2}) below) have been presented in Ref. [\onlinecite{Perez-Piskunow2014}], here we discuss the solution for the full range of parameters and provide more details about its properties, such as the energy dispersion, velocity and chirality of the edge states for both Dirac cones---in particular we analyticaly prove that both cones give rise to states with the same chirality. We also comment on a shortcoming of our solution at the end of this section.

To obtain analytical expressions for the Floquet-edge states within the dynamical gap, we consider a semi-infinite graphene sheet with a zigzag termination. Translational invariance along the edge ($y$ axis) implies that  $u_{mi}(\bm{r})\propto\mathrm{e}^{\ci k_yy}$. Since we are interested in Floquet states that are localized near the edge, we look for solutions of the form $\exp(\kappa x)$ with $\kappa=\ci k_x+q$ and $k_x\,,q\in \Re$. If we take the semi-infinite sheet to be restricted to the $x>0$ region, the physical solution corresponds to  $q<0$---we will keep track of both signs, however, for reasons that will became clear later on.

The boundary condition at the edge of the graphene sheet require for the wavefunction to vanish at one of the lattice sites, say $u_{mB}(x=0)=0$, which in turns requires to combine solutions with $\pm k_x$. After a tedious but straightforward algebra we find that the solutions have the form 
\be
u_{mi}(\bm{r})=\mathcal{C}\, \mathrm{e}^{\ci k_yy} \mathrm{e}^{qx}\mathcal{Q}_{mi}(x)\,,
\label{solA}
\ee
with
\begin{eqnarray}
\nonumber
\mathcal{Q}_{1A}(x)&=& - \sqrt{\frac{1+4\eta^2-\mu^2}{1+\mu}}\, \sin{(k_x x+\theta_k)}\,,\\
\nonumber
\mathcal{Q}_{1B}(x)&=&  \ci \sqrt{1-\mu}\, \sin(k_xx)\,,\\ 
\nonumber
\mathcal{Q}_{0A}(x)&=& \pm\ci\sqrt{1+\mu}\, \sin(k_xx+\varphi_k)\,,\\
\mathcal{Q}_{0B}(x)&=& \mp\sqrt{1+\mu}\,\sin(k_xx)\,,
\end{eqnarray}
where  $\mathcal{C}$ is a normalization constant  and
\begin{eqnarray}
\nonumber
\mathrm{e}^{\ci \varphi_k}&=&\frac{k_y-q+\ci k_x}{|k_y-q+\ci k_x|};
\qquad
\mathrm{e}^{\ci \theta_k}=\frac{k_y+q+\ci k_x}{|k_y+q+\ci k_x|}.\\
\end{eqnarray}
The exponential decay of the wavefunction towards the bulk of the graphene sheet is set by
\begin{equation}
q=\mp \frac{eA_0}{2\hbar c}\sqrt{\frac{1-\mu}{1+\mu}}.
\label{q}
\end{equation}
If we recall that the amplitude of the electric field is $E_0=\Omega A_0/c$, the prefactor in Eq. (\ref{q}) defines half the inverse of the characteristic length 
\begin{equation}
\xi=\frac{\hbar\Omega}{eE_0}\,.
\end{equation}
Hence, the spatial profile of the Floquet topological edge states has a characteristic distance that is \textit{independent} of the graphene's microscopic parameters. This is consistent with the expectation that $\xi$ must be proportional to $\hbar v_F$ divided by the gap, $\xi\sim \hbar v_F/(ev_FA_0/c)=\hbar c/eA_0$. The cancellation of $v_F$ is a consequence of the linear dispersion of graphene.

The oscillating part of the wavefunction towards the bulk of the sample is given by
\begin{eqnarray}
\nonumber
k_x&=&\sqrt{\frac{\varepsilon^2}{(\hbar v_F)^2}-(k_y-q)^2}\\
&=&k_0\sqrt{\frac{1+\mu}{1-\mu}}\sqrt{1-\frac{\mu^2}{\Delta^2}}\,,
\end{eqnarray}
where $\Delta=\eta/\sqrt{1+\eta^2}$ [$\hbar\Omega\Delta$ is the bulk energy gap when calculated with the full $4\times4$ matrix of Eq.~(\ref{matrix_k})].
The corresponding energy dispersion, $\varepsilon(k_y)$,  is obtained from the solution of the following equation
\begin{equation}
\mu(1+\mu)-\eta^2(1-\mu)\mp\eta\, k_y/k_0\sqrt{1-\mu^2}=0\,,\\
\label{disp}
\end{equation} 
which gives two solutions inside the dynamical gap with a real value for $k_x$.
We denote these two solutions as $\phi^A_{K,\mp}(\bm{r})$ to emphasize they correspond to a given Dirac cone ($K$) and to an edge that ends in $A$ atoms--recall, however, that for the chosen $x>0$ region the physical solution corresponds to $q<0$.  The general solution of Eq. (\ref{disp}) can be written in an analytical form but the expression is rather involved to be presented here.
However, close to the middle of the dynamical gap one can approximate the solution as
\begin{equation}
\bep=\frac{\ve}{\hbar\Omega}\approx \frac{(1+2\eta^2)}{2(1+\eta^2)}\pm\frac{\eta}{2(1+\eta^2)}\frac{k_y}{k_0}.
\label{disp1}
\end{equation}
This a linear dispersion, corresponding to massless edge states with a constant velocity (see below).

For $\varepsilon=\hbar\Omega/2$ the solution has a particularly simple form since in that case $k_y=q=-1/2\xi$, $k_x=k_0$ and the wavefunction becomes
\begin{eqnarray}
\label{ana-wave1}
\psi^A_{K,-}(\bm{r},t)&=&\mathrm{e}^{-\ci y/2\xi}\, \mathrm{e}^{-x/2\xi}\, (2 \xi L_y)^{-\frac{1}{2}}\\
\nonumber
&&\times \left[\binom{-\cos{k_0 x} + 2\eta\sin{k_0 x}}{\ci\sin{k_0 x}} \mathrm{e}^{\ci \Omega t}+\binom{\ci \cos{k_0 x}}{-\sin{k_0 x}}\right]
\end{eqnarray}
where we introduced the sample length $L_y$ in the $y$ direction (assume to have periodic boundary conditions).
Note that the oscillation in the direction perpendicular to the edge does not depend on $A_0$ but only on the frequency, $k_0=\Omega/2 v_F$, and that there are many periods in the decay length as  $2k_0\xi=1/\eta\gg1$. 

\begin{figure}[tb]
\includegraphics[width=.95\columnwidth]{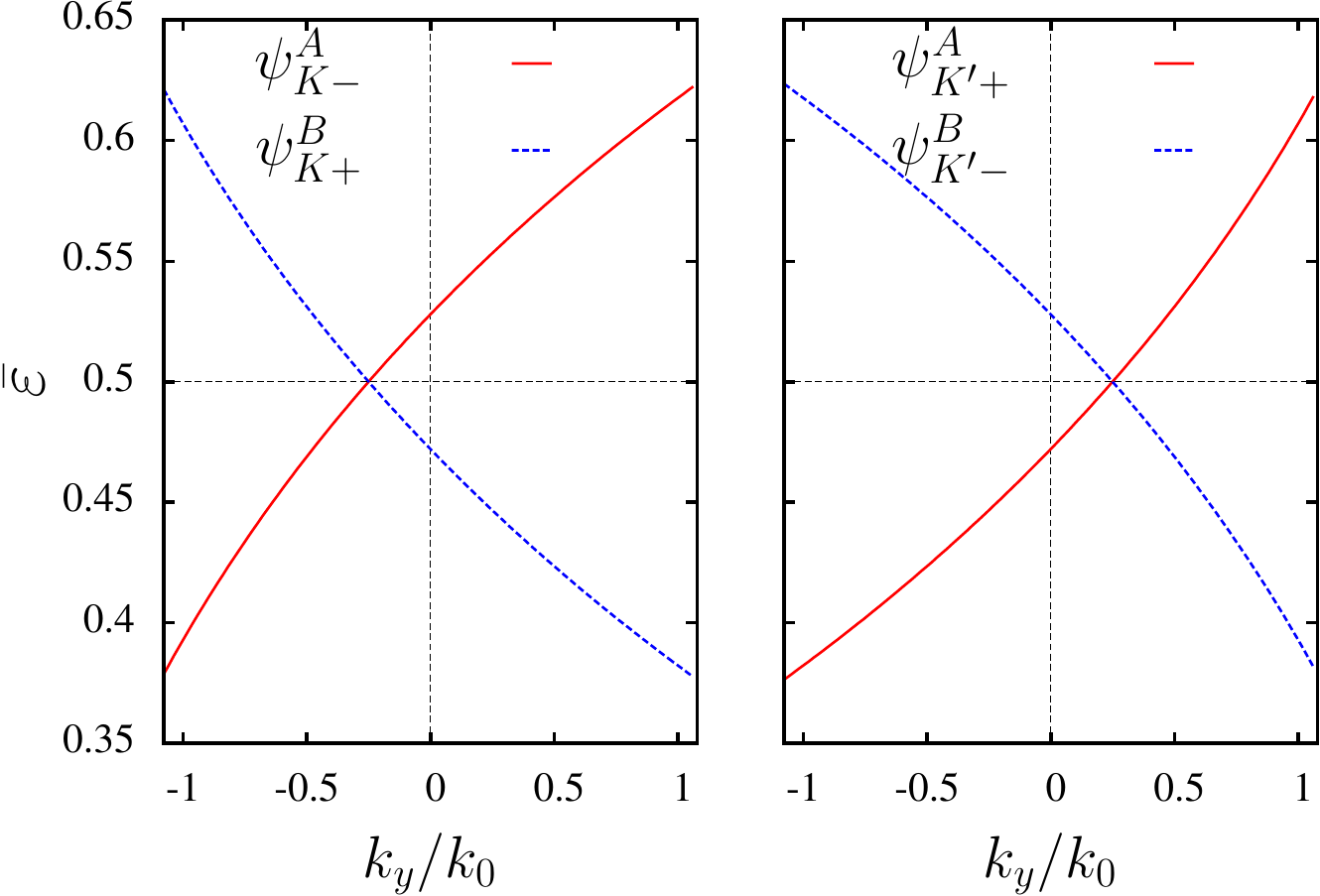}
\caption{Quasi-energy dispersion of the edge states, $\bar{\varepsilon}=\varepsilon/\hbar\Omega$, for the $K$ (a) and $K'$ (b) Dirac cones and $\eta=0.25$. The vertical and horizontal dashed lines indicate the position of Dirac cone and the center of the dynamical gap, respectively. The red (solid) line corresponds to the edge states on a given side of a wide $W\gg\xi$ ribbon while the blue (dashed) one corresponds to the opposite side. Note that the velocity is positive for the former and negative for the latter.}
\label{e_vs_k}
\end{figure}

\begin{figure}[tbp]
\includegraphics[width=\columnwidth]{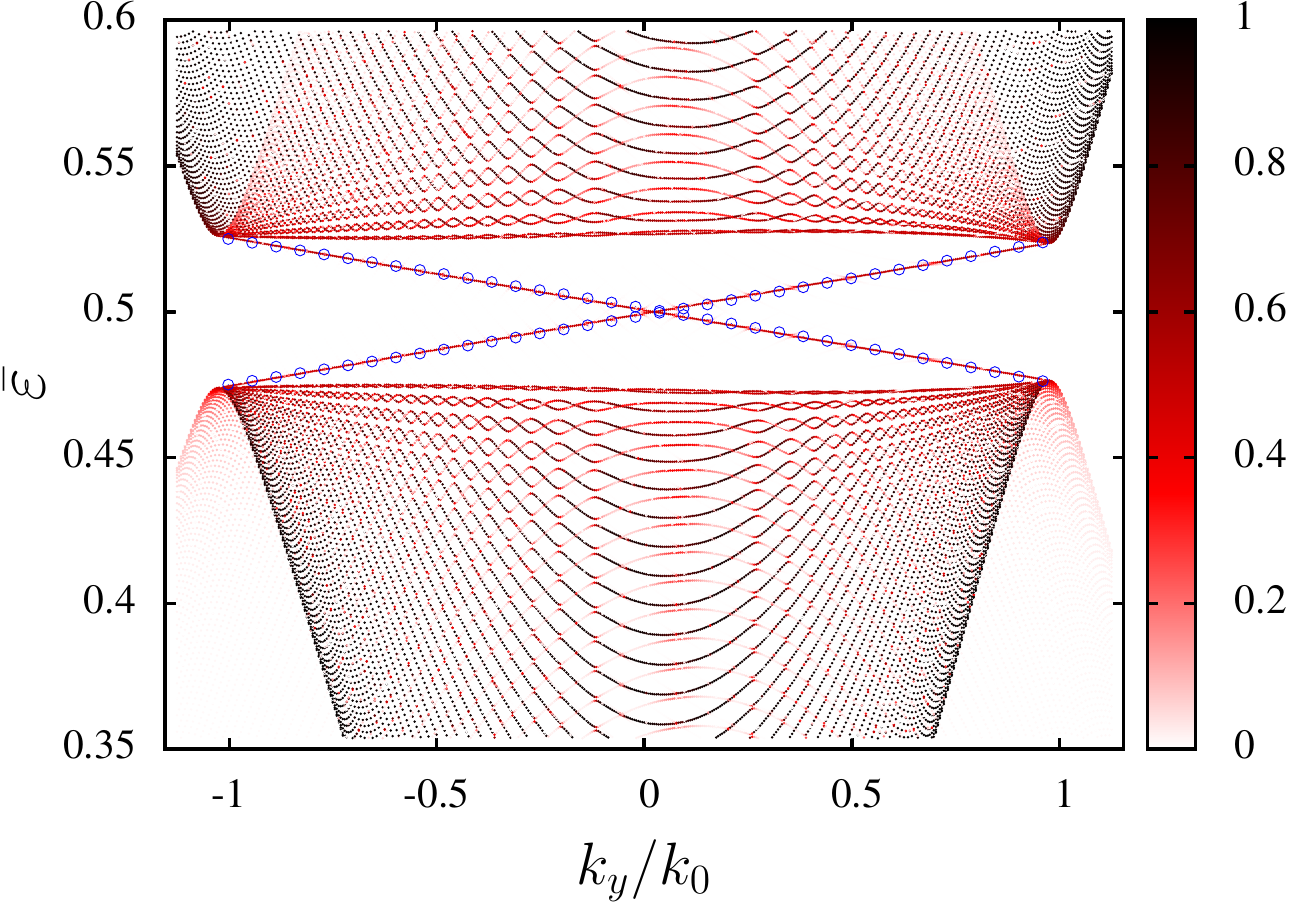}
\caption{(color online) Comparison of the tight-binding quasi-energy dispersion (small points), projected onto the $m=0$ subspace (weight given by the color scale), with the analytical expression (blue open dots) for $h\Omega=0.3\gamma_0$ and $z=10^{-3}$. The  tight-binding data was obtained by numerically solving the Floquet equation with $N_{\mathrm{FR}}=4$ (see main text) and $M=1000$ transverse sites.}
\label{TBdisp}
\end{figure}

For the case of a nanoribbon, we can use this approach to calculate the edge states on the other side of the sample provided the width $W\gg\xi$. In that case, we look for a solution such that $\tilde{u}_{mA}(x=W)=0$.
If we define $\tilde{x}=x-W<0$, we then require an exponential decay $\exp(\tilde{q}\tilde{x})<1$. Hence, the physical solution corresponds to $\tilde{q}>0$, which is consistent with the previous solution---nevertheless, we track both signs of $\tilde{q}$ as before. Following the same procedure we obtain  
\be
\tilde{u}_{mi}(\bm{r})=\mathcal{C}'\, \mathrm{e}^{\ci k_yy} \mathrm{e}^{\tilde{q}\tilde{x}}\tilde{\mathcal{Q}}_{mi}(\tilde{x})
\label{solB}
\ee
with
\begin{eqnarray}
\nonumber
\tilde{\mathcal{Q}}_{1A}(\tilde{x})&=&\sqrt{1-\mu}\, \sin(k_x\tilde{x})\,, \\
\nonumber
\tilde{\mathcal{Q}}_{1B}(\tilde{x})&=& -\ci \sqrt{1-\mu}\, \sin{(k_x \tilde{x}-\theta_k)}\,,\\ 
\nonumber
\tilde{\mathcal{Q}}_{0A}(\tilde{x})&=& \pm\ci \sqrt{1+\mu}\, \sin(k_x\tilde{x})\,,\\
\nonumber
\tilde{\mathcal{Q}}_{0B}(\tilde{x})&=&  \mp  \sqrt{\frac{1+4\eta^2-\mu^2}{1-\mu}}\,\sin(k_x\tilde{x}-\varphi_k)\,,\\
\end{eqnarray}
and 
\begin{equation}
\tilde{q}=\pm\frac{1}{2\xi}\sqrt{\frac{1+\mu}{1-\mu}}\,.
\label{qt}
\end{equation}
Interesting enough, for a given energy, the decay is \textit{di\-ffe\-rent} from the one obtained on the other edge (for a single Dirac cone)---in particular, 
\begin{equation}
q\tilde{q}=-\frac{1}{4\xi^2}\,.
\end{equation}
In addition,
\begin{equation}
k_x=k_0\sqrt{\frac{1-\mu}{1+\mu}}\sqrt{1-\frac{\mu^2}{\Delta^2}}\,,
\end{equation}
while the energy dispersion is obtained from 
\begin{equation}
-\mu(1-\mu)-\eta^2(1+\mu)\mp\eta\, k_y/k_0\sqrt{1-\mu^2}=0\,.\\
\end{equation} 
Note that this solution can be obtained from the previous one by changing $\mu\rightarrow-\mu$. 
Following the previous notation, we denote the corresponding wavefunction as $\phi^B_{K,\pm}(\bm{r})$ . 
Near the middle of the dynamical gap
\begin{equation}
\bep\approx \frac{1}{2(1+\eta^2)}\mp\frac{\eta}{2(1+\eta^2)}\frac{k_y}{k_0}\,,
\label{disp2}
\end{equation}
which again corresponds to massless excitations.
Figure \ref{e_vs_k}(a) shows the Floquet quasienergy dispersion for both solutions, $q<0$ (red solid line) and $\tilde{q}>0$ (blue dashed line), for $\eta=0.25$---a large value to emphasize the symmetries. The symmetry of the spectrum around $\hbar\Omega/2$ is apparent from the figure. The two branches cross at $k_y/k_0=-\eta$. A comparison with a numerical solution of a tight binding model with a larger number of replicas is presented in the next section (Fig. \ref{TBdisp}). The excellent agreement show that our solution correctly describe the system for small values of $\eta$.

For $\varepsilon=\hbar\Omega/2$ we get  $k_y=-\tilde{q}=-1/2\xi$, $k_x=k_0$ and the wavefunction becomes
\bea
\label{ana-wave2}
\psi^B_{K,+}(\bm{r},t)&=& \mathrm{e}^{-\ci y/2\xi}\, \mathrm{e}^{\tilde{x}/2\xi} \frac{1}{\sqrt{2 \xi L_y}} \times\\
\nonumber
&& \left[\binom{\ci \sin{k_0 \tilde{x}}}{- \cos{k_0 \tilde{x}}}\mathrm{e}^{\ci \Omega t}+\binom{\sin{k_0 \tilde{x}}}{-\cos{k_0 \tilde{x}} - 2\eta\sin{k_0 \tilde{x}}}\right]\,.
\eea

The average velocity of the edge states can be readily obtained from the relation $v=(1/\hbar) \partial\ve/\partial k_y$ or, equivalently, by explicitly calculating the average value of the velocity operator $v_F\langle\langle\sigma_y\rangle\rangle$. It is clear from the above figure that the edge states belonging to opposite sides of a finite sample, $\phi^A_{K,-}(\bm{r})$ and $\phi^B_{K,+}(\bm{r})$, have \textit{opposite} velocities. This can be seen explicitly by examining Eqs. (\ref{disp1}) and (\ref{disp2}), from where we find that the velocities are given by 
\begin{equation}
v^A_{K,-}= -v^B_{K,+}=\frac{\eta}{1+\eta^2}  v_F
\end{equation}
near the middle of the gap.

The edge states coming from the other Dirac cone, $K'$, can be obtained from the ones of the $K$ cone if we write the four-component wavefunction as 
$\phi_{K'}(\bm{r})=\{[-\bar{u}_{1B}(\bm{r}),\bar{u}_{1A}(\bm{r})],[-\bar{u}_{0B}(\bm{r}),\bar{u}_{0A}(\bm{r})]\}^\mathrm{T}$ ---such rearrangement is equivalent to apply the operator $-\ci\sigma_y\tau_0$ to the usual four-component wavefunction. In that case, the form of the Hamiltonian for both cones is the same and the physical solutions are $\phi^A_{K',-}(\bm{r})= -\ci\sigma_y\tau_0\phi^B_{K,-}(\bm{r})$ and $\phi^B_{K',+}(\bm{r})= -\ci\sigma_y\tau_0\phi^A_{K,+}(\bm{r})$---here the reason we kept both signs in the previous calculation becomes clear. The corresponding quasi-energy dispersion is shown in Fig. \ref{e_vs_k}(b).  This implies that
\begin{equation}
v^A_{K',-}= v^B_{K,-}\,,\qquad 
v^B_{K',+}=v^A_{K,+}\,.
\end{equation}
Hence, the velocity of the two edge states on a given side of the sample, say $\phi^A_{K,-}(\bm{r})$  and $\phi^A_{K',-}(\bm{r})$, have the same sign. That is, \textit{there are two chiral edge states on each side of the sample}.

Before ending this section it is important to mention a subtle issue regarding the normalization of the wave functions  Eq. (\ref{orto2}). In the notation of  Eq. (\ref{fourier}), the normalization condition implies that 
\begin{equation}
\sum_{m=-\infty}^{\infty} \langle u_m^\alpha| u_{m+n}^\beta\rangle=\delta_{n0}\delta_{\alpha\beta}\,,
\end{equation}
for any integer $n$. This relation is satisfied by the eigenvectors of $\tilde{\mathcal{H}}_F^\infty$  but not necessarily by the ones of the truncated Floquet Hamiltonian $\tilde{\mathcal{H}}_F$ (with any finite number of replicas). This is the case for the solutions shown in Eqs. (\ref{solA}) and (\ref{solB}), except for the important case of $\ve=\hbar\Omega/2$ [Eqs. (\ref{ana-wave1}) and (\ref{ana-wave2})]. That is, the solutions in the $m=0$ and $m=1$ subspaces are not orthogonal in real space ($\la u_1|u_0\ra\ne0$) and so the normalization condition is only satisfied on average and not at all times.
While this is a drawback of our solutions, or any other obtained with a finite number of replicas, it could, in principle, be solved by expanding the solution in powers of the small parameter $\eta$ and incorporating the different orders coming from the different replicas perturbatively. Indeed, if we expand the solutions Eqs.  (\ref{solA}) and (\ref{solB}) to linear order in $\eta$ we can verified that they are correctly normalized at any time to that order.
In any case, one can always compare the approximate analytical solutions with the numerical ones keeping many replicas and check the validity of the former. We do precisely this in the following section.

\section{Atomistic description for laser illuminated graphene}
In this section  we obtain numerical results for the quasi-energy spectrum and the Floquet states  using a tight-binding model for laser illuminated graphene. Our numerical results are shown to compare well with the analytical expressions obtained in the previous section
based on the continuum low-energy model. Moreover, we also explore the laser-induced edge states in: (i) ribbons with terminations other than zigzag, and (ii) a semi-infinite graphene sheet.

An atomistic model for a graphene sheet illuminated by a laser field can be obtained by using a tight-binding Hamiltonian to describe the electrons near the Fermi energy,~\cite{Saito1998,Charlier2007}
\begin{equation}
{\cal H}_g=\sum_{i}\epsilon_{i}^{{}}\,c_{i}%
^{\dagger}c_{i}^{{}}-\sum_{\left\langle i,j\right\rangle }[\gamma_{ij}\,c_{i}%
^{\dagger}c_{j}^{{}}+\mathrm{h.c.}]\,.
\end{equation}
Here $c_{i}^{\dagger}$ and $c_{i}^{{}}$ are the electronic creation and annihilation operators at the $\pi$-orbital on site $i$,  with energy $\epsilon_{i}$, and $\gamma_{ij}$ is the
nearest-neighbors carbon-carbon hopping matrix element, taken to be 
equal to $\gamma_0=2.7$ eV \cite{Dubois2009}. 
The effect of the laser is described through a time-dependent electric field $\bm{E}(t)$ \cite{Oka2009,Savelev2011,Zhou2011}. We choose a
gauge such that $\bm{E}(t)=-(1/c)\, \partial \bm{A}/ \partial t$, where
$\bm{A}$ is the vector potential. In this way, the time dependence of the Hamiltonian enters only through the hopping matrix elements, which acquire a time-dependent phase,\cite{Oka2009,Calvo2012a,Calvo2013}
\begin{equation}
\gamma_{ij}=
\gamma_{0}\exp\left(\mathrm{i}\frac{2\pi}{\Phi_0}\int_{
\bm{r}_i}^{\bm{r}_j}\bm{A}(t)\cdot\mathrm{d}\bm{\ell}\right)\,,
\label{gama}
\end{equation}
where $\Phi_0$ is the magnetic flux quantum.

By using Floquet theory \cite{Camalet2003,Kohler2005,Platero2004,Moskalets2002} as described before one can compute the Floquet spectrum. Once again, one ends up with a time-independent problem in an expanded space. In this case one can picture it as tight-binding problem in a multichannel system where each channel represents the graphene sheet with different number of photons.\cite{Shirley1965, Calvo2013, FoaTorres2014} The Floquet Hamiltonian has the same structure as in Eq. (\ref{matrix_inf}) where the Dirac Hamiltonian is replaced by $\mathcal{H}_g$ and the coupling between replicas is changed accordingly. It is worth mentioning that in tight-binding method the time dependent perturbation is never purely harmonic given the exponential dependence of Eq. (\ref{gama}) on the radiation field amplitude. Hence, there is a coupling among all the replicas\cite{Calvo2013} and not just those with $\Delta m=\pm1$---nevertheless, for $\eta\ll 1$, only the later are relevant. The results of the continuous model are recovered if the dimensionless parameter $z=2\pi a_{cc} A_0/\Phi_0$, is  much smaller than unity.\cite{notes5} Here $a_{cc}$ is the carbon-carbon distance. 
In terms of this parameter, the relevant magnitudes can be written as $\eta=(3\gamma_0/2\hbar\Omega) z$
and $\xi =a_{cc}/z$.
%

\begin{figure}[tbp]
\includegraphics[width=\columnwidth]{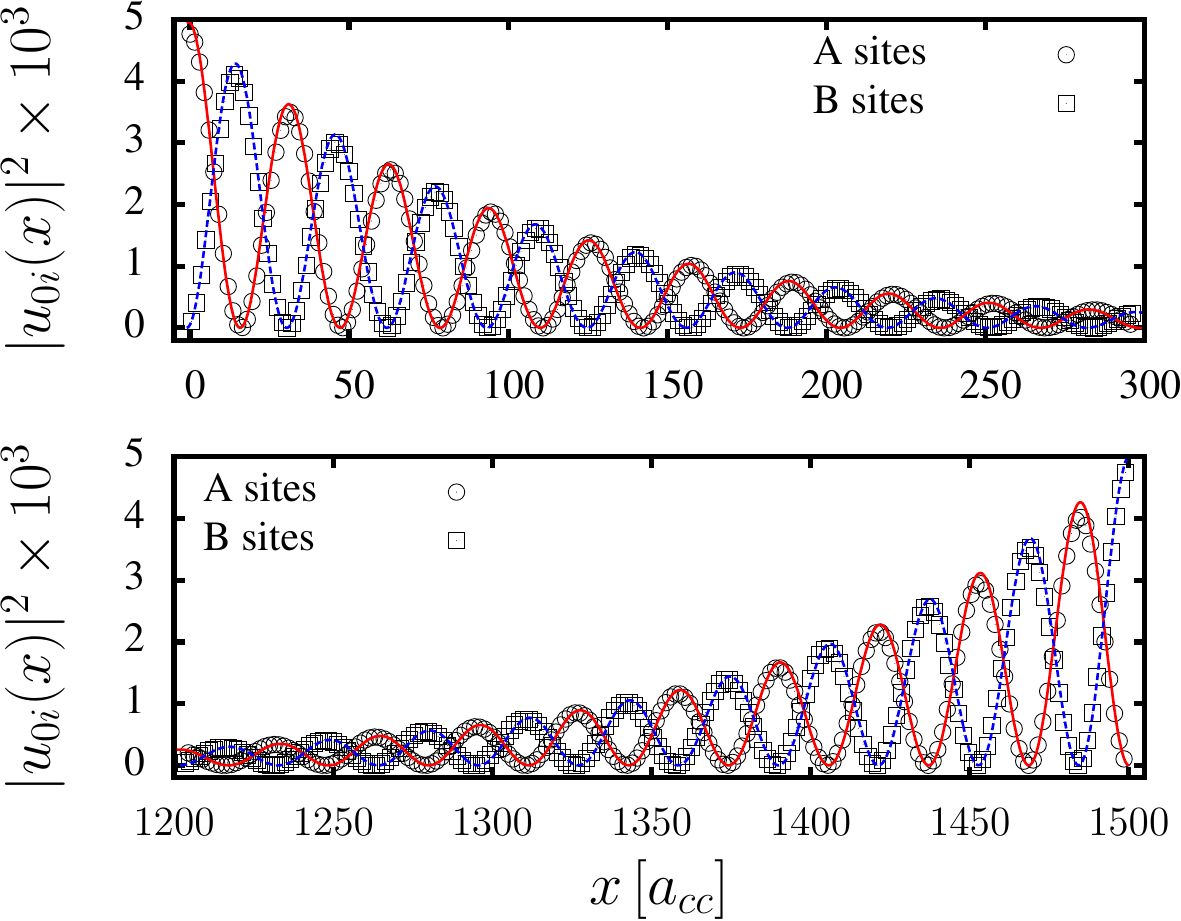}
\caption{(color online) Comparison of the modulus squared of numerically obtained wave function (symbols) projected onto the $m=0$ subspace with the analytical expression (solid and dashed lines). The empty circles (empty squares) correspond to the numerical results for the A-sites (B-sites) while the analytical expressions are shown with solid and dashed lines, respectively. These results were computed for a laser with $h\Omega=0.3\gamma_0$ and  $z=0.01$. The plotted wavefunctions correspond to the two branches with $\varepsilon\sim0.15 \gamma_0$ and $N_\mathrm{F}=4$.}
\label{compar-wave}
\end{figure}

\subsection{Comparison between analytical and numerical results for a finite ribbon}
\label{comparison}

The tight-binding model can be solved for a ribbon of finite width ($M$ being the number of transverse sites) or for a semi-infinite sheet. We deal with the former case in this subsection. To such end we obtain the Floquet spectrum, and the corresponding wavefunctions, by numerical diagonalization of the Floquet Hamiltonian on the Bloch basis 
\be
\Ht_{gF}^k=\Ht_{gF}^\mathrm{uc}+V \mathrm{e}^{\ci k_yd}+V^\dagger \mathrm{e}^{-\ci k_yd}\,.
\ee
Here $\Ht_{gF}^\mathrm{uc}$ is the Floquet Hamiltonian corresponding to one unit cell (transverse layer), $V$ is the hopping matrix between unit cells, $d$ is distance between them and $k_y$ the Bloch wavevector along the ribbon. The size of this matrix is $M\times N_\mathrm{FR}$, which imposes a limitation on both size ($M$) and  number of Floquet replicas ($N_\mathrm{FR}$)---we typically used $M\lesssim 2000$ and $N_\mathrm{FR}\lesssim 6$. 

\begin{figure}[t]
\includegraphics[width=\columnwidth]{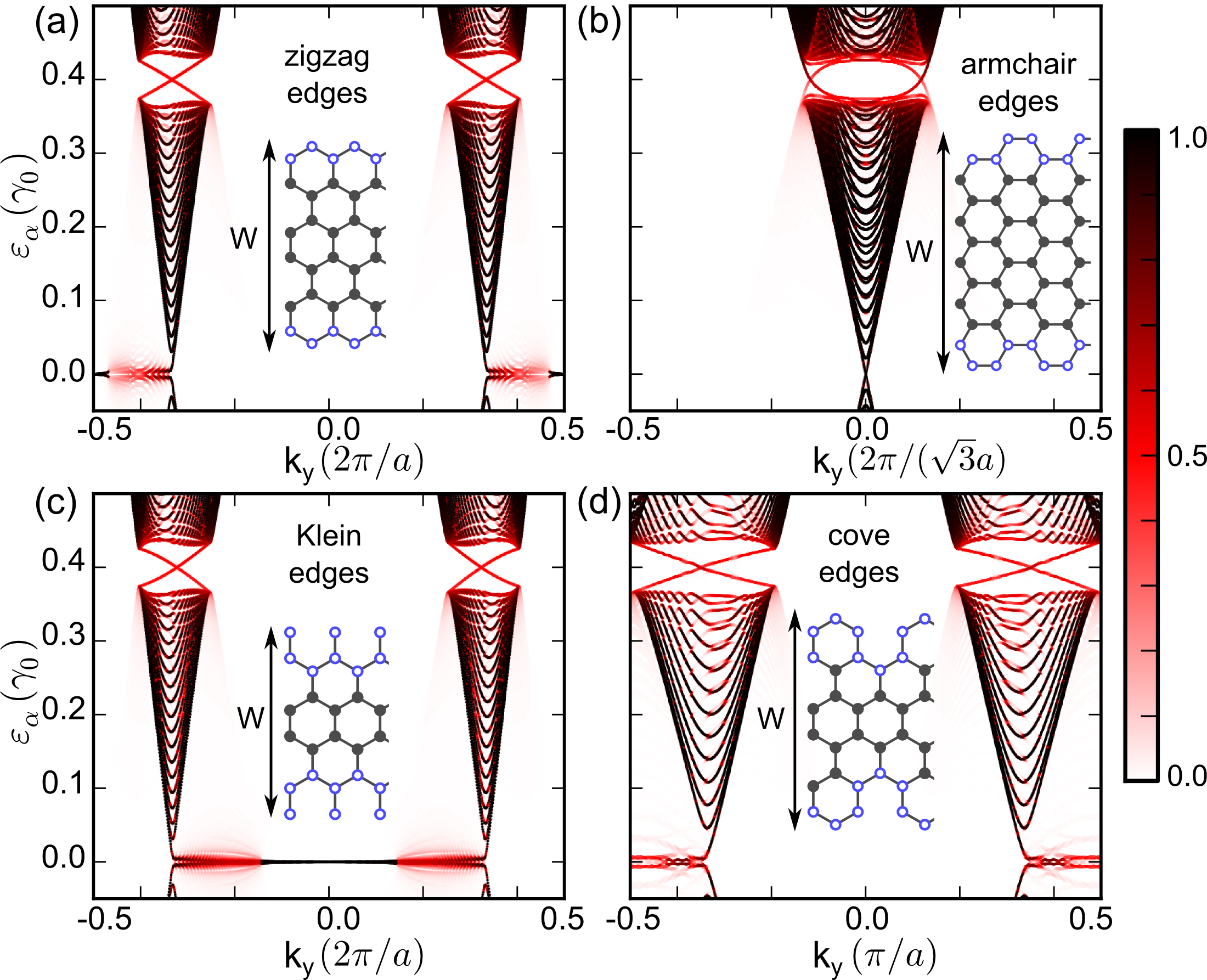}
\caption{(color online) Quasi-energy spectrum for different ribbon terminations: (a) zigzag, (b) armchair, (c) zigzag with Klein edges and (d) zigzag with cove edges. The color scale indicates the weight contributing to the average density of states. In the calculation the ribbons are being illuminated with circularly polarized light with $\hbar\Omega=0.8\gamma_0$ and $z=0.04$, panels a, c and d are for ribbons with $W=31.95$ nm while panel b has $W=31.48$ nm. All calculations include the Floquet replicas between $n=-2$ and $n=+2$. The laser-induced states bridging the bulk dynamical gap are evident while those at the Dirac point are hardly developed.}
\label{other_edge}
\end{figure}

A comparison of the results, that includes the $m=2,1,0,-1$ replicas  ($N_\mathrm{FR}=4$), with the analytical solution is shown in Fig. \ref{TBdisp} for a finite width zigzag ribbon ($M=1000$ transverse sites). The agreement is very good in the entire gap, despite the fact that the tight-binding calculation shows signatures of trigonal warping for the chosen energies.

Figure \ref{compar-wave} shows a comparison of the wave function for $\varepsilon=0.1503 \gamma_0$ obtained with the tight-binding method and the analytical result (Eqs. (\ref{ana-wave1}) and (\ref{ana-wave2})) for a $212$nm wide ribbon--the other parameters are indicated in the caption of the figure. Each panel of the fi\-gu\-re corresponds to one of the two branches of a given Dirac cone. Note that they are located at opposite sides of the ribbon.

Besides zigzag ribbons we have also tested the emergence of laser-induced edge states for other ribbon terminations. Figure \ref{other_edge} shows a few typical cases: armchair ribbons (Fig. \ref{other_edge}-b), zigzag nanoribbons with Klein edges\cite{Suenaga2010} (Fig. \ref{other_edge}-c) and cove edges\cite{Wakabayashi2010} (Fig.~\ref{other_edge}-d). The case of a zigzag ribbon is also shown for comparison in Fig. \ref{other_edge}-a. In all cases laser-induced edge states bridging the dynamical gap do emerge. In contrast, the smallness of the gap at the Dirac point together with the finite system size conspire against the formation of the edge states at the Dirac point, which are hardly developed for the parameters used in Fig.\ref{other_edge}.
Moreover, one can observe  that the quasi-energy dispersion of the laser-induced edge states close to the Dirac point is much more sensitive to the edge type. While for zigzag and zigzag-like edges (Fig.\ref{other_edge} a, c, d) the laser slightly bends the natural occurring flat bands, for armchair edges the bands crossing at the Dirac point remain while the others retire away from the Dirac point, thereby forming the bulk gap.

\subsection{Laser-induced edge states in a semi-infinite graphene sheet}
\label{semi-infinite}

For the semi-infinite case we use the recursive Green function method to obtain the local Floquet-Green's functions near the edge of a very wide ribbon ($M\approx 2^{20}-2^{25}$, and, eventually, a large $N_\mathrm{FR}$). Namely, we calculate $G^0_{jj}(\ve+\ci0^+,k)=\bra{j,0}[(\ve+\ci0^+)I-\Ht_{gF}^k]^{-1}\ket{j,0}$, where $\ket{j,0}$ represents the state on the $j$-transverse site on the $m=0$ replica.  This is an extremely efficient method that allow us to obtain, among other quantities, the time-averaged local spectral function\cite{Oka2009,Zhou2011}
\begin{equation}
{\rho}_{jj}(\varepsilon,k)= -\frac{1}{\pi} 
\mathrm{Im}[G^0_{jj}(\ve+\ci 0^+,k)]\,.
\label{Floquet-DOS}
\end{equation}
This way, we can visualize the quasi-energy dispersion by plotting the density of states near the edge,
\be
\rho_\mathrm{edge}(\varepsilon,k)= \sum_{j<r} {\rho}_{jj}(\varepsilon,k)\,,
\ee
where $r$ is choose to satisfy $r\gg\xi/a_{cc}$ in order to capture the total weight of the edge states. 
\begin{figure}[t]
\includegraphics[width=\columnwidth,clip=true]{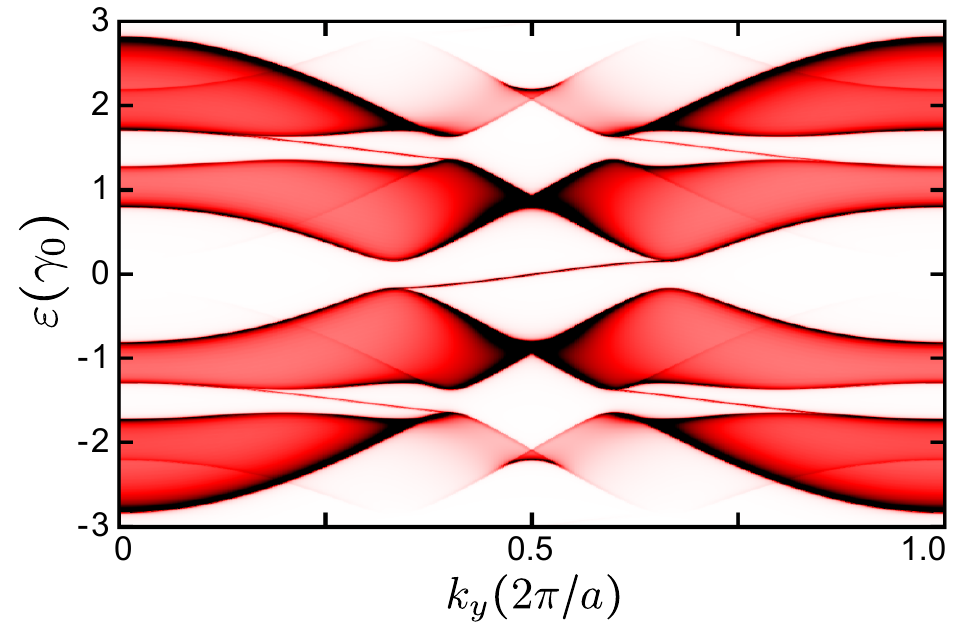}
\caption{(color online) Color map of the averaged local density of states, $\rho_\mathrm{edge}(\varepsilon,k)$,  projected onto several sites near the edge of the semi-infinite sheet, $r=50$, as a function of $k_y$ and $\ve$. Here, $\hbar\Omega=3\gamma_0$, $z=0.5$, $a=\sqrt{3} a_{cc}$ and $N_\mathrm{FR}=5$. The appearance of edge states bridging the gaps is apparent from the figure. Notice that the two states at the dynamical gap have the opposite chirality than the one appearing inside the gap developed at the Dirac point. }
\label{Gfull_big}
\end{figure}

Figure \ref{Gfull_big}  present the results for $\hbar\Omega=3\gamma_0$ and $z=0.5$, that corresponds to $\xi=2a_{cc}$ (so we took $r=50$). Here we used very large parameters for the radiation fields which are unrealistic for graphene, but that allow us to make a few important points: (a) there is only one chiral edge state, on each valley, that bridges the dynamical gap as we are looking at one edge of a semi-infinite sheet; (b) there is a single edge state at the Dirac point with the \textit{opposite} chirality ($\partial \ve/\partial k_y$ is negative at the dynamical gap and positive at the Dirac point).
The extended dark areas on the plot correspond to the bulk states---normalization of the color scale is done for presentation purposes only.
 
\begin{figure}[ptb]
\includegraphics[width=\columnwidth]{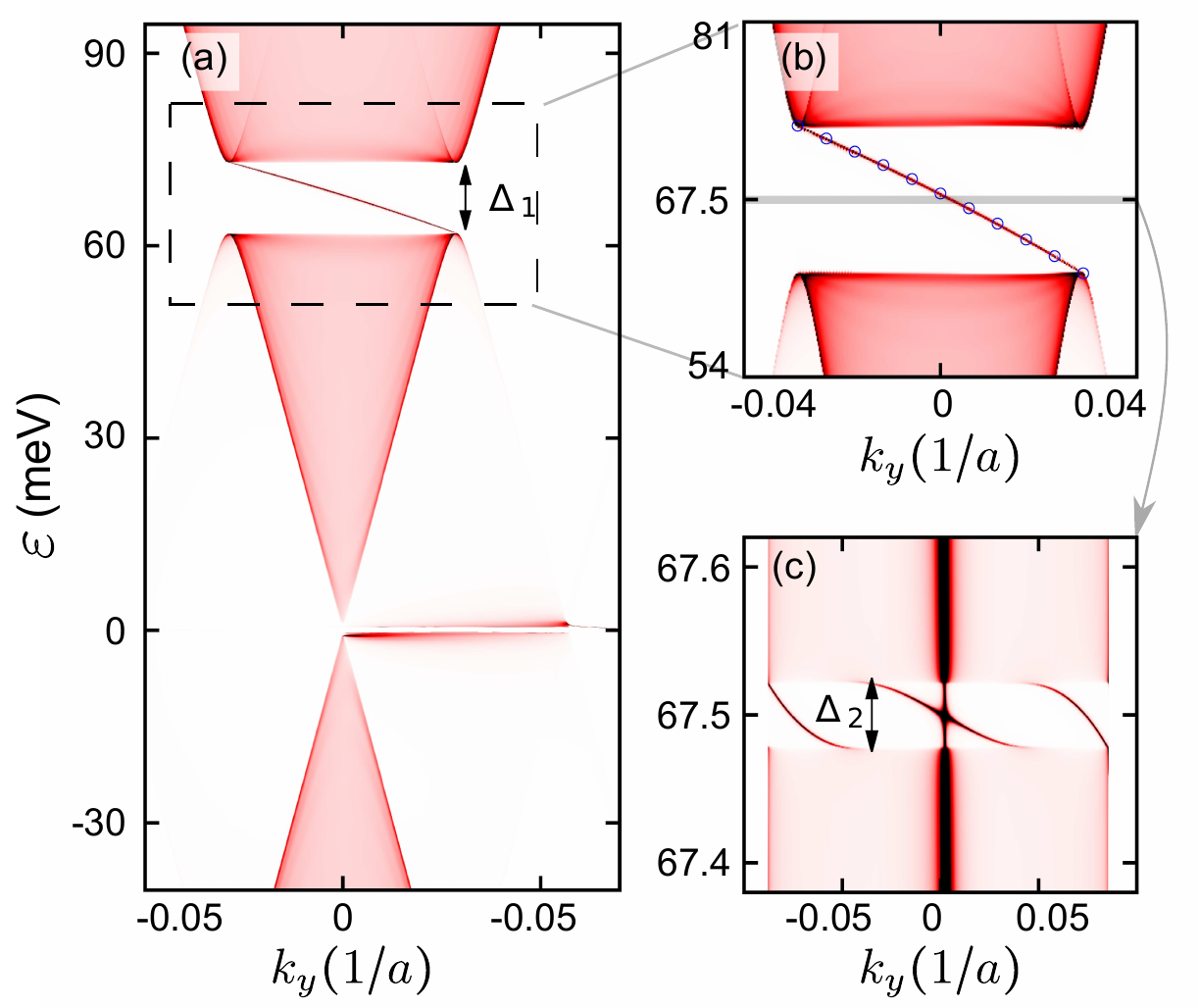}
\caption{(color online) Color map of $\rho_\mathrm{edge}(\varepsilon,k)$ for a realistic set of radiation parameters---color scale is set for visualisation purposes (a) Large energy span showing both the dynamical $\hbar\Omega/2$ gap ($\Delta_1\sim\eta\hbar\Omega$) and the Dirac point gap with they corresponding edge states. (b) Zoom in on the $\hbar\Omega/2$ gap; dots correspond to the analytical solution. (c) Further zoom in near the middle of the \textit{first} order gap showing the emergence of a second generation gap  ($\Delta_2\sim\eta^3\hbar\Omega$) and the corresponding edge states. Note that the energy (momentum) scale is reduced (enlarged). Inside this gap there are three additional states---one of them leads to an anticrossing with the first order edge state.}
\label{Grealista}
\end{figure}

In contrast to the finite size case, where one has to tune the parameters of the radiation field so that $\xi$ is several times smaller than the ribbon width, otherwise the edge channels mix and split,\cite{Perez-Piskunow2014} the Green's function method imposes essentially no limit for the radiation intensity. This is shown in Fig.~\ref{Grealista}(a) where we plot $\rho_\mathrm{edge}(\varepsilon,k)$ for a realistic mid-infrared field in graphene: $\hbar\Omega=0.05\gamma_0$ ($135$meV), $z=2.8\times10^{-3}$. We included here a large number of replicas, $N_\mathrm{FR}=16$, although there is essentially no different in the results if we use, say, $N_\mathrm{FR}=6$ (not shown).  
The  large energy span allows to see both the gap at $\hbar\Omega/2$ and  at  the Dirac point. The later is narrower and the corresponding edge state is less developed. In particular, near $k_y=0$, the mixing with the $m=\pm1$ replicas is strong enough as to 
completely blurred it, becoming more clearly resolved only  beyond $k_y=2k_0$ where the above mentioned replicas have no states (the mixing with higher order replicas is not discernible on this scale). 
This is yet another indication of the weakness of the edge state at the Dirac point as compared with the ones that occur at the dynamical gap.

A closer view of the dynamical gap ($\hbar\Omega/2$), is shown in Fig.~\ref{Grealista}(b).
The absence of finite size effects allows for a clear development of the edge states, in agreement with the analytical results (indicated by the open dots).

At this point it is worth mentioning a subtle point that is usually overlooked when discussing Floquet edge states: the effect of the mixing with high order replicas ($m\geq2$ and $m\leq-1$) on the edge state that occurs inside the \textit{first order} dynamical gap. This effect is hardly visible for the realistic parameters used in Fig.~\ref{Grealista}, but a zoom in allows to detect such anomalies: Fig.~\ref{Grealista}(c) shows the development of a second generation gap with additional edge states. The appearance of a hierarchy of `gaps' deserves some clarification. The first order gap ($\Delta_1\sim \eta \hbar \Omega$) arises from the mixing of the $m=0$ and $m=1$ replicas. As we shown in the previous section it contains \text{one} edge states (per valley). As the next order replicas are included, $m=-1$ and $m=2$, the first order gap is partially filled and a second generation gap develops inside it of order $\Delta_2\sim\eta^3 \hbar\Omega$. In this case, the first order edge state adquieres some broadening (Fig.~\ref{Grealista}(c)) and, related to this,  a parametrically small extended component of the wave function. In addition, three additional edge states appear inside the second generation gap, given a total of four edge states. This scheme continues upon adding more and more replicas leading to a further reduction of the actual gap of the Floquet spectrum.
While in the continuos Dirac-like approximation [Eq. (\ref{matrix_inf})] there is never a true  gap, there are always higher order replicas that contribute to the density of states at any quasi-energy and thus close it, in the tight binding model there is always a gap since replicas with $\delta m\gtrsim \mathcal{O}(D/\hbar\Omega)$ do not overlap.\cite{Rudner2013} Here $D$ is the bandwidth of graphene.  The actual energy gap in the tight binding model, however, is much smaller than the first order gap of Eq. (\ref{autoval}), roughly $\eta^{D/\hbar\Omega} \hbar\Omega$. It is only inside this later gap that a true topologically protected states exist. Nevertheless, when $\eta\ll1$, the contribution  to $\rho_\mathrm{edge}(\varepsilon,k)$ decays exponentially with the number of replicas,  so that the average density of states is dominated by the first order effect as shown in Fig.~\ref{Grealista}. If we are interested in dc properties, such as dc currents, the average density is what matters for identifying the dominant contributions. 

While the high order edge states are beyond the scope of the present work since they are not relevant for realistic implementations in graphene---a detailed analysis  will be presented elsewhere---we would like to briefly  mention the following. The number of edge states $N$ appearing inside a given gap depends on the number of Floquet replicas considered in the calculation. The rule is (for a given valley)
\be
N=\sum_i |\delta m_i|\,,
\ee
where $\delta m_i$ is the difference of the Floquet indexes of the pairs of replicas that become degenerated at the dynamical gap,  leading to a high order gap of order $\eta^{|\delta m_i|}$, and $i$ runs over the replicas retained. For instance, the first generation gap contains $1=|1-0|$ edge states, the second one contains $4=|1-0|+|2-(-1)|$, the third contains  $9=|1-0|+|2-(-1)|+|3-(-2)|$ and so on---we have checked this for the mentioned generations. This result is valid as far as the continuum approximation remains  a good description of the band and can be obtained by constructing an effective $2\times2$ Hamiltonian that describes the crossing between each pair of replicas (see appendix). Notice that $\delta m_i$ also corresponds to the number of photon's processes involved. What we would like to stress is that while this number of edge states is what a calculation of the Chern number would give, converging only when $\mathcal{O}(D/\hbar\Omega)$ replicas are retained, \textit{only the first order state gives a significant contribution to the averaged density of states}. Hence, caution should be taken when deducing transport properties from the Chern numbers of the Floquet Hamiltonian alone.

\section{Summary and Conclusions}

We focused the emergence of Floquet edge states in irradiated graphene by using complementary approaches: a simple analysis of the topological character of the bulk Floquet bands based on a continuum model, an explicit analytical solution for the states developing at the dynamical gaps and numerical calculations. The topological arguments contain a discussion of a few novel aspects: (i) the analysis close to the dynamical gaps which suggests a topological phase which is different from the one at the gap close to the Dirac point (one has two chiral edge states bridging the gap and the other just one); (ii) a discussion on the relevance of different Floquet replicas for the calculation of Chern numbers, here we argue that a Floquet-projected calculation (on the $m=0$ channel) captures the physics of time-averaged magnitudes such as the dc density of states.

On the other hand, the analytical solutions provide valuable information such as the scaling of the decay length of these Floquet chiral edge states with the system parameters, which is not easily accesible neither through a bulk calculation nor numerical simulations and that could serve as a guide to experiments. Those results are complemented with numerics for different ribbon terminations, highlighting the generality of the physics described for zigzag ribbons. Further insight is also provided by a numerical calculation for a semi-infinite graphene sheet. This allows to discuss subtle issues that are hard to access when considering a finite width.

All these results highlight the experimental accessibility of the edge states at the dynamical gap
 in graphene as opposed to the one found at the Dirac point. The former offer also the possibility to tune the transport energy window where they appear.
As for the experimental signature of these Floquet edge states, one can anticipate\cite{Foa2014} the appearance of a Hall-like voltage in a radiated graphene sample whenever the Fermi energy of the reservoirs lines up with the dynamical gap. This voltage should change sign if the circular polarization is reversed from, say, clockwise to counterclockwise. A Hall signal should also develop at the Dirac point\cite{Kitagawa2011} but, interestingly enough, with the opposite sign.

\section{Acknowledgements}
We acknowledge financial support from PICTs 2008-2236, 2011-1552 and
Bicentenario 2010-1060 from ANPCyT, PIP 11220080101821 and 11220110100832 from CONICET and 06/C415 SeCyT-UNC. GU and LEFFT acknowledge support from the ICTP associateship program, GU also thanks the Simons Foundation.
\appendix

\section{Chern number calculation of the Floquet bands}

In the  time-independent case, the topology of a system can be characterized by the Chern numbers associated to each one of the Bloch bands. Namely,
\bea
\nonumber
C_n&=&\frac{\ci}{2\pi}\oint_\mathcal{C}\, \bra{u_{n\bm{k}}}\bm{\nabla}_{\bm{k}}\ket{u_{n\bm{k}}}\cdot d\bm{k}\\
&=&\frac{1}{\pi} \mathrm{Im}\int_{\mathrm{BZ}}\,  \la\partial_{k_y}u_{n\bm{k}}|\partial_{k_x}u_{n\bm{k}}\ra\,d^2k \,,
\eea
where $n$ is the band index, $\ket{u_{n\bm{k}}}$ is the periodic part of the Bloch eigenfunction, $\mathcal{C}$ the contour of the Brilluoin zone. Alternatively, the later expression can be cast in the following form
\be
C_n=\frac{1}{2\pi}\int_{\mathrm{BZ}}\,  \bm{\Gamma}_{n\bm{k}}\cdot d\bm{S}_{\bm{k}} \,,
\ee
with
\bea
\nonumber
\bm{\Gamma}_{n\bm{k}}&=&\mathrm{Im}\sum_{m\ne n}\frac{\bra{u_{n\bm{k}}}\bm{\nabla}_{\bm{k}}H_{\bm{k}}\ket{u_{m\bm{k}}}\times\bra{u_{m\bm{k}}}\bm{\nabla}_{\bm{k}}H_{\bm{k}}\ket{u_{n\bm{k}}}}{(\ve_{n\bm{k}}-\ve_{m\bm{k}})^2}\,.\\
\eea
The latter expression makes evident that  the main contribution to $C_n$ comes from the points in the Brillouin zone near an avoided crossing, that is, where the gap between the $n$ band and the nearest bands is small. 

In our case, we can apply the same procedure to the bulk Floquet Hamiltonian to characterize the topological properties of the Floquet bands and the corresponding edge states.\cite{Rudner2013}
While a direct calculation using the above expression with the full tight binding Hamiltonian is possible, though numerically rather demanding if $\hbar\Omega\ll D$---recall that $\mathcal{O}(D/\hbar\Omega)$ replicas are required to include all anti crossings---,we used the continuos model, and some further approximations, to gain some insight.

The dynamical gap at  $\hbar\Omega/2$ occurs  in $k$-space near a point where, in the absence of radiation,  there is a degeneracy between a pair of replicas at that energy for the same value of $k$. Such degeneracies appear at $k_p=(2p+1)k_0$ with $p$ an integer number. That is, the $m=0$ and $m=1$ replicas become degenerated at $k=k_0$, the $m=2$ and the $m=-1$ at $k=3k_0$ and so on.
Since, as we pointed out above, the Chern number is dominated by the contribution near the degeneracy points, in order to get the contribution coming from a given region it is sufficient to obtain a $2\times2$ effective Hamiltonian, valid for $k\sim k_p$. In that case, by writing it as 
$\mathcal{H}^{\mathrm{eff}}_F(\bm{k},p)=\bm{h}_p(\bm{k})\cdot\bm{\sigma}$, one can obtain the contribution to the Chern by calculating\cite{Hasan2010}
\be
c_p=\frac{1}{4\pi}\int\, \hat{\bm{h}}_p\cdot\left(\partial_{k_x}\hat{\bm{h}}_p\times\partial_{k_y}\hat{\bm{h}}_p\right)\, d^2k\,.
\label{cp}
\ee

Following this procedure, and explicitly calculating $\bm{h}_p(\bm{k})$, we found that the number of edge states $N$ appearing inside the dynamical gap depends on the number of Floquet replicas and it is given by
\be
N=\sum_p c_p=\sum_i |\delta m_i|
\ee
where $\delta m_i$ is the difference of the Floquet indexes of the pairs of replicas that become degenerated at the dynamical gap,  leading to a high order gap of order $\eta^{|\delta m_i|}$, and $i$ runs over the replicas retained. It is worth emphasising that $N$ becomes independent of the number of replicas only when  $\mathcal{O}(D/\hbar\Omega)$ are included and that $|\delta m_i|$ corresponds to the number of photons involved in the process that couple the corresponding replicas.


\begin{thebibliography}{69}
\bibitem [{not({\natexlab{a}})}]{notes3}%
  \BibitemOpen
  \href@noop {} {} \bibinfo {note} {A general solution will
  be of the form $\ket{\phi_{\alpha}(t)}=\sum_\alpha a_\alpha
  \exp(-\ci\varepsilon_{\alpha}t/\hbar)\ket{\psi_{\alpha}(t)}$, with $a_\alpha$
  time-independent coefficients.}\BibitemShut {Stop}%
\bibitem [{not({\natexlab{b}})}]{notes4}%
  \BibitemOpen
  \href@noop {} {} \bibinfo {note} {Strictly speaking, there
  is no gap in the Floquet Hamiltonian in the continuum model if the infinite
  replicas are kept. However, any realistic model with a finite bandwidth will
  always have a finite gap. This gap, however, will be much smaller that the
  first order gap calculated here. Nevertheless, the contribution of the edge
  states to the transport properties, will be mainly dominated by the first
  order gap.}\BibitemShut {Stop}%
\bibitem [{not({\natexlab{c}})}]{notes5}%
  \BibitemOpen
  \href@noop {} {} \bibinfo {note} {Note that a strong laser
  intensity may also result in higher order effects like photon-assisted
  tunneling through the dynamical gap.}\BibitemShut {Stop}%
\end{thebibliography}

%

\end{document}